\documentclass[aps,10pt,pra,superscriptaddress,showpacs,floatfix,longbibliography,notitlepage]{revtex4-1}
\usepackage{dsfont}
\usepackage{cprotect}
\usepackage{graphicx}
\usepackage{amsmath}
\usepackage{amssymb}
\usepackage{hyperref}
\usepackage{bm}
\usepackage{subfigure}
\hypersetup{colorlinks=true,linkcolor=cyan,citecolor=magenta,filecolor=magenta,urlcolor=cyan,runcolor=cyan}
\usepackage[pdftex]{color}
\newcommand{\e}{\text{e}}
\newcommand{\beq}{\begin{equation}}
\newcommand{\eeq}{\end{equation}}
\newcommand{\beqnn}{\begin{equation*}}
\newcommand{\eeqnn}{\end{equation*}}
\newcommand{\bea}{\begin{eqnarray}}
\newcommand{\eea}{\end{eqnarray}}
\newcommand{\beann}{\begin{eqnarray*}}
\newcommand{\eeann}{\end{eqnarray*}}
\newcommand{\bes} {\begin{subequations}}
\newcommand{\ees} {\end{subequations}}
\newcommand\tr{{\mbox{Tr\,}}}

\begin{document}
\title{Calculating the divided differences of the exponential function \\ by addition and removal of inputs}

\author{Lalit Gupta}
\affiliation{Department of Physics and Astronomy and Center for Quantum Information Science \& Technology, University of Southern California, Los Angeles, California 90089, USA}

\author{Lev Barash}
\affiliation{Landau Institute for Theoretical Physics, 142432 Chernogolovka, Russia}
\affiliation{Information Sciences Institute, University of Southern California, Marina del Rey, California 90292, USA}

\author{Itay Hen}
\email{itayhen@isi.edu}
\affiliation{Department of Physics and Astronomy and Center for Quantum Information Science \& Technology, University of Southern California, Los Angeles, California 90089, USA}
\affiliation{Information Sciences Institute, University of Southern California, Marina del Rey, California 90292, USA}

\begin{abstract}
\noindent We introduce a method for calculating the divided differences of the exponential function by means of addition and removal of items from the input list to the function. Our technique exploits a new identity related to divided differences recently derived by F. Zivcovich [Dolomites Research Notes on Approximation 12, 28-42 (2019)]. We show that upon adding an item to or removing an item from the input list of an already evaluated exponential, the re-evaluation of the divided differences can be  
done with only $O(s n)$ floating point operations and $O(s n)$ bytes of memory, where $[z_0,\dots,z_n]$ are the inputs and $s \propto \max_{i,j} |z_i - z_j|$. We demonstrate our algorithm's ability to deal with input lists that are orders-of-magnitude longer than the maximal capacities of the current state-of-the-art.  We discuss in detail one practical application of our method: the efficient calculation of weights in the off-diagonal series expansion quantum Monte Carlo algorithm.
\end{abstract}
\maketitle

\section{Introduction: Divided differences of the exponential function}
The divided differences~\cite{Thomson1933,dd:67,deboor:05} of a function $f(\cdot)$ with respect to real- (or complex-)valued inputs $[z_0,\ldots,z_n]$ is defined as
\beq\label{eq:divideddifference2}
f[z_0,\ldots,z_n] \equiv \sum_{j=0}^{n} \frac{f(z_j)}{\prod_{k \neq j}(z_j-z_k)} \,.
\eeq
The above expression is ill-defined if two (or more) of the inputs are repeated, in which case $f[z_0,\ldots,z_n]$ can be properly evaluated using limits~\footnote{As will be discussed later on, other definitions of divided differences exist, which do not necessitate the use of limits in the case of repeated indices~\cite{Opitz1964,deboor:05}.}. For instance, in the case where $z_0=z_1=\ldots=z_n=x$, the definition of divided differences reduces to: 
\beq
f[x,x,\ldots,x] = \lim_{z_i \to x}f[z_0,\ldots,z_n] = \frac{f^{(n)}(x)}{n!} \,,
\eeq 
where $f^{(n)}(\cdot)$ stands for the $n$-th derivative of $f(\cdot)$. 

Divided differences have historically been used for computing tables of logarithms and trigonometric functions, for calculating the coefficients in the interpolation polynomial in the Newton form and more.
A central question related to divided differences has to do with the computational cost of accurately evaluating 
the divided difference of the exponential function (DDEF) $f(z)=\e^z$
as a function of number of inputs~\cite{Mccurdy1984,Higham2011,Zivcovich2019}, 
namely, the evaluation of $\exp[z_0,\ldots,z_n]$ with increasing $n$.

DDEFs can be calculated in a straightforward manner via the recurrence relation
\bea\label{eq:ddr}
f[z_i,\ldots,z_{i+j}] =\frac{f[z_{i+1},\ldots , z_{i+j}] - f[z_i,\ldots , z_{i+j-1}]}{z_{i+j}-z_i} \,,\quad i=0,\ldots,n-j,\quad j=1,\ldots,n,
\eea 
with the initial conditions $f[z_i] = f(z_{i})$, $i= 0,\ldots,q$ (assuming that the points are distinct).
In practice however, such a straightforward approach is known to produce imprecise and numerically unstable results if simple double-precision floating point arithmetic is used~\cite{Mccurdy1984,Higham2002accuracy,Higham2011,Caliari2007,Zivcovich2019}.
To overcome these precision issues, several methods for the accurate evaluation of DDEFs have been devised over the years. 
These have employed an identity known now as Opitz's formula~\cite{Opitz1964,Mccurdy1980} which gives various divided differences in terms of functions of matrices, explicitly:
\beq
\label{eq:Opitz}
f\begin{pmatrix} 
z_0 & 1    &    &  \\
      & z_1 & 1 &   \\
&& \ddots & \ddots \\
&&& z_{n-1} & 1 \\
&&&& z_n 
\end{pmatrix}
=
\begin{pmatrix} 
f[z_0] & f[z_0,z_1] & \dots & f[z_0,z_1,\dots,z_n] \\
&f[z_1] & \dots & f[z_1,\dots,z_n] \\
&& \ddots & \vdots \\
&&& f[z_n]
\end{pmatrix},
\eeq
where the $(1,n+1)$-th entry of the output matrix on the right-hand side is the desired quantity. 
A hybrid algorithm for the accurate computation of DDEFs that uses the above formula was suggested by McCurdy \emph{et al.}~\cite{Mccurdy1984} which prescribes the standard recurrence relations when it is safe to do so, and otherwise employs a Taylor series expansion of the exponential
in conjunction with Opitz's theorem and repeated matrix scaling-and-squaring~\cite{doi:10.1137/04061101X}.

Observing that only a single entry of the matrix Eq.~(\ref{eq:Opitz}) is actually needed to obtain the DDEF, Opitz's formula was later used to estimate DDEFs via the calculation of the action of the matrix exponential on a given vector~\cite{Higham2011,Higham2002accuracy,Higham2008functions,Guttel} employing
$O(n)$ matrix-vector products, or $O(n^3)$ floating-point operations in the general case.
However, while these general-purpose routines for the computation of the matrix exponential 
are accurate in terms of the matrix norm,
they do not ensure high accuracy for individual matrix elements (which is needed for the accurate evaluation of divided differences).

Improving upon the above, in a recent work by F. Zivcovich~\cite{Zivcovich2019}, a new identity related to divided differences was derived
which allows for a faster and more accurate computation of
the vector of DDEFs
\beq
\label{eq:vectorf}
\mathbf{f} = (\exp[z_0], \exp[z_0, z_1], \dots, \exp[z_0,\dots, z_n])^T\,,
\eeq
requiring $O(sn^2)$ floating point operations and $O(n^2)$ bytes of memory, where
$s=\lceil \frac{1}{3.5} \max_{i,j} |z_i - z_j| \rceil$. 

In this study, we propose an algorithm, based on Zivcovich's identity, for a fast and accurate calculation of 
DDEFs by means of sequential addition or removal of items from the input list to the function, which allows us to update the DDEF with changes to the input list. 
With each addition or removal, the recalculation of the DDEF is performed using only $O(s n)$ floating
point operations and $O(s n)$ bytes of memory. As we show, our method enables the calculation of DDEFs of essentially unlimited input sizes without the need to resort to resource-demanding and time-consuming arbitrary-precision arithmetic.
Our primary motivation for devising the algorithm is the need for such a routine in the recently introduced off-diagonal 
series quantum Monte Carlo (QMC)~\cite{ODE,ODE2,PMR} which we discuss in detail later on. 

The paper is organized as follows. We begin with briefly reviewing Zivcovich's derivations insofar as they are relevant to our purposes in Sec.~\ref{sec:ZivcovichReview}.
We discuss our method in Secs.~\ref{sec:add}-\ref{sec:precision} and in Sec.~\ref{sec:bench} we benchmark and analyze the algorithm's performance.
We demonstrate the applicability of our method to the calculation of configuration weights in quantum Monte Carlo simulations of quantum many-body systems in Sec.~\ref{sec:QMC} and discuss the significance of our results as well as an outlook in Sec.~\ref{sec:conclusions}.

\section{Calculating DDEFs by addition and removal of inputs}
\label{sec:method}
\subsection{Zivcovich's  algorithm for evaluating DDEFs}
\label{sec:ZivcovichReview}

We begin our discussion describing Zivcovich's algorithm~\cite{Zivcovich2019} for computing the vector $\mathbf{f}$, Eq.~(\ref{eq:vectorf}),
introducing along the way several modifications which allow us (among other things) to obtain Opitz's matrix of divided differences column by column, rather than row by row as in the original algorithm.
The main building block of  Zivcovich's algorithm is obtaining the vector
\beq
\label{eq:hj}
\mathbf{h}^{(j)}=\left(\exp\left[\frac{z_j}{s}\right],\exp\left[\frac{z_j}{s},\frac{z_{j-1}}{s}\right]s^{-1},\dots,\exp\left[\frac{z_j}{s},\dots,\frac{z_{j-N}}{s}\right]s^{-N}\right)^T
\eeq
from the vector
\beq
\mathbf{h}^{(j-1)}=\left(\exp\left[\frac{z_{j-1}}{s}\right],\exp\left[\frac{z_{j-1}}{s},\frac{z_{j-2}}{s}\right]s^{-1},\dots,\exp\left[\frac{z_{j-1}}{s},\dots,\frac{z_{j-1-N}}{s}\right]s^{-N}\right)^T
\eeq
using only $O(n)$ operations, with the initial condition $\mathbf{h}^{(-1)}=\left(1,(1!\cdot s)^{-1},(2!\cdot s^2)^{-1},\dots,(N!\cdot s^N)^{-1}\right)^T$. Here, the vectors $\mathbf{h}^{(j)}$ have length $N\geq n+30$ ($z_k=0$ for every $k<0$).
Zivcovich observed that the matrix operation that takes $\mathbf{h}^{(j-1)}$ to $\mathbf{h}^{(j)}$ can be written as 
\beq
\label{eq:transformcol}
H_1(z_j-z_{j-1})H_2(z_j-z_{j-2})\dots H_N(z_j-z_{j-N})\,,
\eeq
where $H_i(z)$ is the $(N+1) \times (N+1)$ identity matrix, with the value $z$ added at the $(i,i+1)$-th position, i.e.,
$$H_i(z) = \begin{pmatrix} \mathbf{I}\left(i-1\right)&&&\\ &1&z&\\&&1\\&&&\mathbf{I}\left(N-i\right) \end{pmatrix},$$
with $\mathbf{I}(i)$ denoting the $i \times i$ identity matrix.
Each operator $H_i(z)$ can be applied with only $O(1)$ operations.
Applying the transformation Eq.~(\ref{eq:transformcol}) successively for $j=0,1,\dots, n$, one obtains the matrix $G=RF(z_0/s,z_1/s,\dots,z_n/s)R^{-1}$, where
\beq
F(z_0,z_1,\dots,z_n) = 
\begin{pmatrix} 
\exp[z_0] & \exp[z_0,z_1] & \dots & \exp[z_0,z_1,\dots,z_n] \\
&\exp[z_1] & \dots & \exp[z_1,\dots,z_n] \\
&& \ddots & \vdots \\
&&& \exp[z_n]
\end{pmatrix}
\eeq
is the divided differences matrix and $R=\text{diag}\{1,s,\ldots,s^n\}$. The above procedure gives the columns of $G$ successively from left to right. The matrix $F$ is obtained from $G$ via the relation
\beq
\label{eq:scaling}
F(z_0,z_1,\dots,z_n) = R F\left(z_0/s,z_1/s,\dots,z_n/s\right)^s R^{-1} = G\left(z_0/s,z_1/s,\dots,z_n/s\right)^s \,,
\eeq
which follows from Opitz's formula~\cite{Mccurdy1984} and
gives the first row of $F(z_0,z_1,\dots,z_n)$
as $\mathbf{f}^T = \mathbf{g}^{(s)}$, where $\mathbf{g}^{(1)}$ 
is the first row of the matrix $G$,
and $\mathbf{g}^{(i+1)}=\mathbf{g}^{(i)}G$ for $i=1,2\dots,s-1$.

The conditions $N\geq n+30$ and $s\geq\lceil \frac{1}{3.5} \max_i |z_i| \rceil$ ensure that the first $n+1$ elements of the vector Eq.~(\ref{eq:hj}) are calculated to a machine-level accuracy, which is in turn a sufficient condition for the accurate calculation of
the matrix $G$. The above conditions follow from the fact that a double precision accuracy in the calculation of $\e^z$
is attainable by truncating its Taylor series at order $n_*=30$ provided that $|z|\leq z_*=3.5$~\cite{Zivcovich2019,Higham2011,Caliari2018} 
(in passing, we note that choices other than $(n_*=30, z_*=3.5)$ may exist which may even provide potentially more rapidly converging routines~\cite{Higham2002accuracy,Higham2011}).

In addition, that $\exp[z_0,z_1,\dots,z_n]$ can be written as $e^{\mu} \exp[z_0 - \mu,z_1-\mu,\dots,z_n-\mu]$,
allows us to choose any scaling parameter $s$ greater than \hbox{$s_{\mathrm{\min}}(\mu,n) = \lceil \frac{1}{3.5} \max_{i} |z_i-\mu| \rceil$}
as long as $\mu$ is subtracted from all inputs and the end result is multiplied by $\e^\mu$.
The shifting by $\mu$ may translate to a substantial reduction of the scaling parameter $s$ whenever
$s_{\mathrm{\min}}(\mu,n)<s_{\mathrm{\min}}(0,n)$.
Choosing $\mu$ to be the mean $\bar{z}=\frac1{n+1}\sum_{i=0}^n z_i$ 
ensures that $s=\lceil \frac{1}{3.5} \max_{i,j} |z_i-z_j| \rceil$
is a suitable choice for the scaling parameter because it is larger 
than $s_{\mathrm{\min}}(\mu,n)$~\cite{Zivcovich2019}.

Having reviewed Zivcovich's algorithm, we are now in a position to discuss our algorithm for computing\break $\exp{[z_0,\ldots,z_{n-1},z_n]}$ given that the DDEF with 
$[z_0,\ldots,z_{n-1}]$ has already been calculated. In what follows, we shall treat the input list to the algorithm as a stack and discuss the re-evaluation of the DDEF under addition of items to the stack and removal of items from it.
We will further demonstrate how our algorithm solves in an efficient manner one limitation shared by all existing algorithms to date, which has to do with the precision of the calculation at large values of $n$ and $s$.
Our main result is that the re-evaluation associated with the addition to and removal from the input stack can be done with only $O(s n)$ floating point operations and $O(s n)$ bytes of memory for arbitrarily long input lists.

\subsection{Adding an item to the input stack\label{sec:add}}

We first consider the case where the DDEF vector $(\exp[z_0], \exp[z_0, z_1], \dots, \exp[z_0,\dots, z_{n-1}])^T$
has already been calculated for some $n>0$,
and that the vector  $\mathbf{h}^{(n-1)}$ and the rows $\mathbf{g}^{(i)}$, $i=1,\dots,s$, as defined above, are stored in memory.
The addition of a new input item $z_{n}$ (assuming for the time being that the addition does not 
require increasing the scaling parameter $s\ge s_{\mathrm{\min}}$) is carried out as follows.
The upper triangular matrix $G(z_0/s,\dots,z_{n}/s)$ should contain the additional last column 
$\mathbf{w}^{(n)}=(s^{-n}\exp[z_0/s,z_1/s,\dots,z_n/s],s^{n-1}\exp[z_1/s,\dots,z_n/s],\dots,s^0\exp[z_n/s])^T$ 
when compared to the already stored $G(z_0/s,\dots,z_{n-1}/s)$
(other additional elements are zeros at positions $(n+1,1), \dots, (n+1,n)$ ).
All elements of the new column are contained in the vector $\mathbf{h}^{(n)}$ defined in Eq.~(\ref{eq:hj}) with $j=n$.
Applying the transformation Eq.~(\ref{eq:transformcol}) with $j=n$ on the already-stored $\mathbf{h}^{(n-1)}$, requires $O(n)$ operations:
\beq
\label{eq:transformcol2}
\mathbf{h}^{(n)}=H_1(z_n-z_{n-1})H_2(z_n-z_{n-2})\dots H_N(z_n-z_{n-N})\mathbf{h}^{(n-1)}.
\eeq

The row $\mathbf{g}^{(1)}$, which coincides with the first row of the matrix $G(z_0/s,\dots,z_{n}/s)$, 
contains an additional item
$\exp[z_0/s,\dots,z_{n}/s]/s^{n}$ as compared to the first row of $G(z_0/s,\dots,z_{n-1}/s)$.
This item too can be found in $\mathbf{h}^{(n)}$.
Each of the rows $\mathbf{g}^{(i)}$, $i=2,\dots,s$, should be appended an element
$\mathbf{g}^{(i-1)}\mathbf{w}^{(n)}$. In particular, the required 
last element of $\mathbf{f}^T = \mathbf{g}^{(s)}$ is obtained as $\mathbf{g}^{(s-1)}\mathbf{w}^{(n)}$.
Storing the vector $\mathbf{h}^{(n)}$ and the rows $\mathbf{g}^{(i)}$, $i=1,\dots,s$, requires only $O(s n)$ bytes of memory.
Thus, our addition routine requires only $O(s n)$ operations and $O(s n)$ bytes of memory.

In the initial case of $n=0$, i.e., when the first input is added to the empty list,
the vector $\mathbf{h}^{(0)}$ is obtained from $\mathbf{h}^{(-1)}$ employing Eq.~(\ref{eq:transformcol2}), and each of the rows $\mathbf{g}^{(i)}$ consists of a single element
$\mathbf{g}^{(i)}_0=\left(h^{(0)}_0\right)^i$, $i=1,\dots,s$,
where $h^{(0)}_0 = \e^{z_0/s}$ as per Eq.~(\ref{eq:hj}).

In addition to the above procedure, all input values can offset by a fixed amount $\mu$ 
(where the final DDEF value is multiplied by $\e^\mu$),
in order to obtain a smaller scaling parameter $s$.
Both values $\mu$ and $s$ should be set at the initial stage and cannot be changed
during the subsequent additions. Nonetheless,
the scaling parameter $s$ should not exceed
$s_{\mathrm{\min}}(\mu,n)$ after every addition.
To ensure that, we choose initially $\mu=z_0$ if $z_0$ is the only known value
at first.
If several values $z_0,z_1,\dots,z_k$ are known initially, 
then the optimal choice of $\mu$ is the mean $\frac1{k+1}\sum_{i=0}^k z_i$ in which case a proper choice for the scaling parameter $s$
is $s=\lceil \frac{1}{3.5} \max_{i,j} |z_i-z_j| \rceil$,
which is larger than $s_{\mathrm{\min}}$.
Whenever the addition of a new item $z_n$ requires a larger value of $s$, which happens when $s_{\mathrm{\min}}(\mu,n)$ exceeds $s$, the algorithm restarts and the matrix $G$ is recalculated from scratch by sequentially adding all of $z_0,\ldots,z_n$ again with the updated values of $\mu$ and $s$.
Importantly, we find that in order to ensure the high accuracy of the resulting values, the value $N$ should remain unchanged during the updates
to the vector of divided differences. Since having $N\geq n+30$ is also important for maintaining accuracy,  each time $n$ exceeds $N-30$ we double $N$, reallocate memory for the vector $\mathbf{h}$
and for the rows $\mathbf{g}^{(i)}$, $i=1,\dots, s$ and recalculate the vector of divided differences. 

\subsection{Removing an item from the input stack}
\label{sec:remove}

Removing an item from the top of the input stack, an operation that is of importance in applications where DDEF inputs need to be replaced or updated, can be done with relative ease. 
Item removal requires applying the inverse of the transformation Eq.~(\ref{eq:transformcol2}), which produces 
$\mathbf{h}^{(n-1)}$ from $\mathbf{h}^{(n)}$, explicitly:
\beq
\label{eq:transformcol3}
\mathbf{h}^{(n-1)} = H_N(z_{n-N}-z_n) H_{N-1}(z_{n-N+1}-z_n)\dots H_1(z_{n-1}-z_n) \mathbf{h}^{(n)}.
\eeq
Item removal also necessitates the removal of the last element from each of the rows $\mathbf{g}^{(i)}$, $i=1,\dots,s$ including the last item of the DDEF vector
$\mathbf{f} = \left(\mathbf{g}^{(s)}\right)^T$.
We thus find that DDEF re-evaluation following the removal of the last item $z_n$ requires only $O(n)$ floating point operations.

\subsection{Direct computation of the modified DDEFs with higher precision}
\label{sec:rescale}

A natural limitation of the above algorithms stems from the fact that DDEFs decrease rapidly with $n$.
This follows directly from the bounds~\cite{Farwig1985}
\beq
e^{\overline{z}} \le n!\cdot \exp[z_0,z_1,\dots,z_n] \le \overline{e^z},
\eeq
where $\overline{z}=\frac1{n+1}\sum_{i=0}^n z_i$
and $\overline{e^z}=\frac1{n+1}\sum_{i=0}^n e^{z_i}$.
Since $\exp[z_0,z_1,\dots,z_n]=e^{\overline{z}} \exp[z_0 - \overline{z},z_1-\overline{z},\dots,z_n-\overline{z}]$,
one may assume without loss of generality that $\overline{z}=0$, in which case the value of $\exp[z_0,z_1,\dots,z_n]$ scales as $1/n!$.
This implies that computing DDEFs for large $n$ using standard double-precision arithmetic is not feasible.

We overcome the above obstacle by choosing to compute, rather than $\mathbf{f}$ given in Eq.~(\ref{eq:vectorf}),  the modified vector 
\beq
\label{eq:tildef}
\mathbf{\tilde f} = \left(\exp[z_0], 1!\cdot \exp[z_0, z_1], \dots, n!\cdot \exp[z_0,\dots, z_n]\right)^T
\,.
\eeq
The range of the elements in $\mathbf{\tilde f}$ is substantially narrower than that of $\mathbf{f}$;
the elements of the modified vector $\mathbf{\tilde f}$ are not smaller than 1 when $\overline{z}=0$,
and as such can be computed more precisely than elements of $\mathbf{f}$. To obtain the vector $\mathbf{\tilde f}$,
the following modifications to the algorithm described in Sec.~\ref{sec:ZivcovichReview} are in order.

Rather than transforming the vectors $\mathbf{ h}^{(j)}$ one must now transform the modified vectors $\mathbf{\tilde h}^{(j)}$ defined as:
\beq
{\tilde h}^{(j)}_i = \left\{
\begin{array}{ll}                                
h_i^{(j)}\cdot j!/(j-i)! & \quad\mbox{for } i=0,1,\dots,j\\
h_i^{(j)}\cdot i! & \quad\mbox{for } i=j+1,\dots,N.
\end{array}\right.
\eeq
In addition, the vector $\mathbf{\tilde h}^{(-1)}$ should be initialized to $\mathbf{\tilde h}^{(-1)}=(1,s^{-1},\dots,s^{-N})^T$.
Furthermore, rather than constructing the matrix $G$ above, we construct the matrix $\tilde G$ which is related to $G$ via $\tilde G_{ij}=G_{ij}\cdot j!/i!$. The rows $\mathbf{\tilde g^{(i)}}$ are defined as $\tilde g^{(i)}_j = g^{(i)}_j \cdot j!$, where $i=1,2,\dots, s$ and $j=0,1,\dots,n$.

The above rescaling also necessitates the use of the modified transformation
$\tilde H_1^{(j)}(z_j-z_{j-1})\tilde H_2^{(j)}(z_j-z_{j-2})\dots \tilde H_N^{(j)}(z_j-z_{j-N})$
in lieu of Eq.~(\ref{eq:transformcol}), where
\beann
\tilde H_i^{(j)}(z) &=& \begin{pmatrix} \mathbf{I}\left(i-1\right)&&&\\ &j/(j-i+1)&z/(j-i+1)&\\&&1\\&&&\mathbf{I}\left(N-i\right) \end{pmatrix},
\qquad 1\leq i\leq j,\\
\tilde H_i^{(j)}(z) &=& \begin{pmatrix} \mathbf{I}\left(i-1\right)&&&\\ &1&z/i&\\&&1\\&&&\mathbf{I}\left(N-i\right) \end{pmatrix},
\qquad i>j.
\eeann
Similarly, $\tilde H_i^{(n)}$ are to be used instead of $H_i$ in Eqs.~(\ref{eq:transformcol2}) and (\ref{eq:transformcol3}).

We note that in the special case of $s=1$, only the first row of the matrix $\tilde G$ is needed, so the values $\tilde h_i^{(j)}$ for $i=0,1,\dots,j-1$ and $j=0,1,\dots,n$, which may require the use of extended precision data types, are not used in the modified DDEF calculation.
In this case, we find that the input list sizes for which modified DDEFs can be accurately calculated using the modified transformations is practically unlimited.
Numerical tests confirm that employing the modified relations allows us, in the $s=1$ case, to obtain accurate values of the vector $\mathbf{\tilde f}$
using standard double-precision floating point arithmetic at least up to $n \approx 10^5$.
This is to be contrasted with the original formulation of Sec.~\ref{sec:ZivcovichReview}, 
in which case the DDEF values create an underflow already at $n \approx 100$. We demonstrate the above in Sec.~\ref{sec:bench}.

\cprotect\subsection{Accurate computation of DDEFS for large $n$ and $s$}
\label{sec:precision}
An additional limiting factor of the algorithm (shared by all other algorithms as well) is that the usage of double-precision floating point data types to evaluate DDEFs is a source of inaccuracies when $n$ or $s$ are too large. This stems from the (relatively) limited range of the double-precision floating point data type $\sim[10^{-308},10^{308}]$. 

The above limitation can however be overcome if floating-point data structures with extended-precision are used instead. Arbitrary precision data types are however expected to be too costly and will slow down the algorithm considerably, as the memory requirements as well as the runtimes of basic arithmetic operations for these grow with $n$ and $s$~\cite{MPFR}. This extra cost can nonetheless be avoided if one observes that the additional precision is not required for the mantissa of the floating point but rather only for its exponent. 
The range of the mantissa can be shown to be sufficient by considering the relative error $\varepsilon$ of each floating point operation.
After $N_f$ floating point operations, the overall accumulated error may be approximated at $\varepsilon \sqrt{N_f}$. For a double precision floating point we have $\varepsilon \approx 10^{-17}$ and so after $N_f = 10^{22}$ operations the overall error is only on the order of $10^{-6}$.

Taking the above into account, we devise a new floating-point data structure, that we refer to as \verb#ExExFloat# (an abbreviation of {\it extended exponent floating point}) 
using which, calculations with values in the extended range
$\sim [10^{-646456992},10^{646456992}]$ are possible.
The \verb#ExExFloat# consists of a mantissa, which in itself is a double-precision floating point, and an additional $32$-bit integer for its exponent. 
Arithmetic operations such as addition, multiplication and division with this new data type are performed using only several elementary operations on double-precision floats. 
A \verb#c++# realization of our algorithm including an implementation of the \verb#ExExFloat# data type is available on GitHub~\cite{DivDiffCode}.

\section{Numerical testing\label{sec:bench}}

We next present the results of several benchmarking tests carried out to verify the computational complexity of the routines devised above. We calculate execution runtimes of the addition and removal routines described in Secs.~\ref{sec:add} and~\ref{sec:remove} on randomly generated input lists of different sizes $n$ and different scaling parameters $s$. Our algorithms are coded in \verb#c++# and compiled with a \verb#g++# compiler with a \verb#-O3# optimization. 
Benchmarking was done on an Intel Core i5-8257U CPU running at $1.4$GHz.

\cprotect\subsection{Calculation of DDEFs with extended precision versus double precision floating point data structures}
\label{sec:bench:exexfloat}

Here, we examine the behavior of calculation times with $n$ and $s$, comparing the performance of our algorithm against implementations that do not use the improved formulation introduced in Sec.~\ref{sec:rescale} or the specially devised extended precision data type discussed in Sec.~\ref{sec:precision}. 

\begin{figure}[t]
\includegraphics[width=0.482\textwidth]{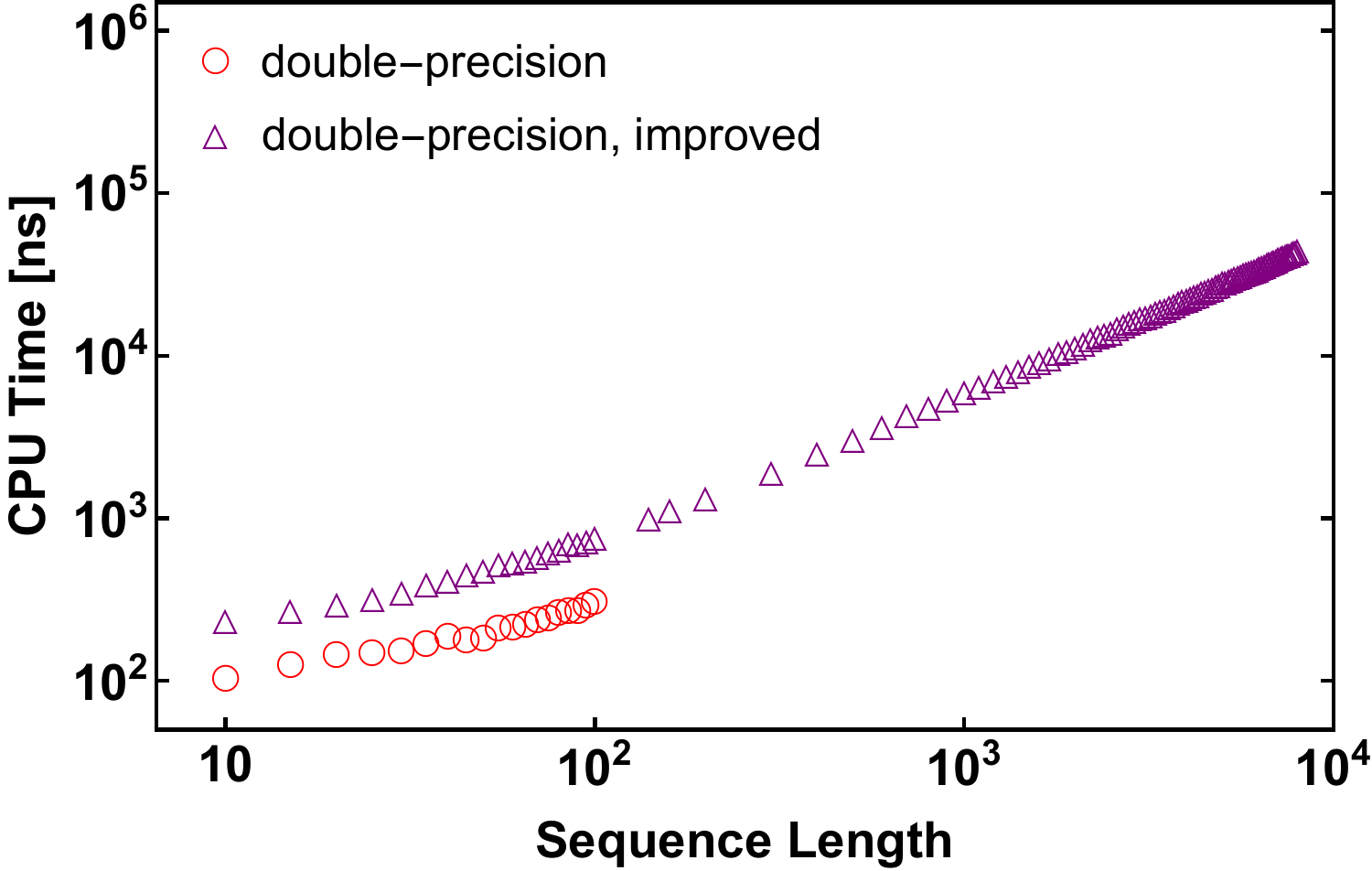}
\includegraphics[width=0.482\textwidth]{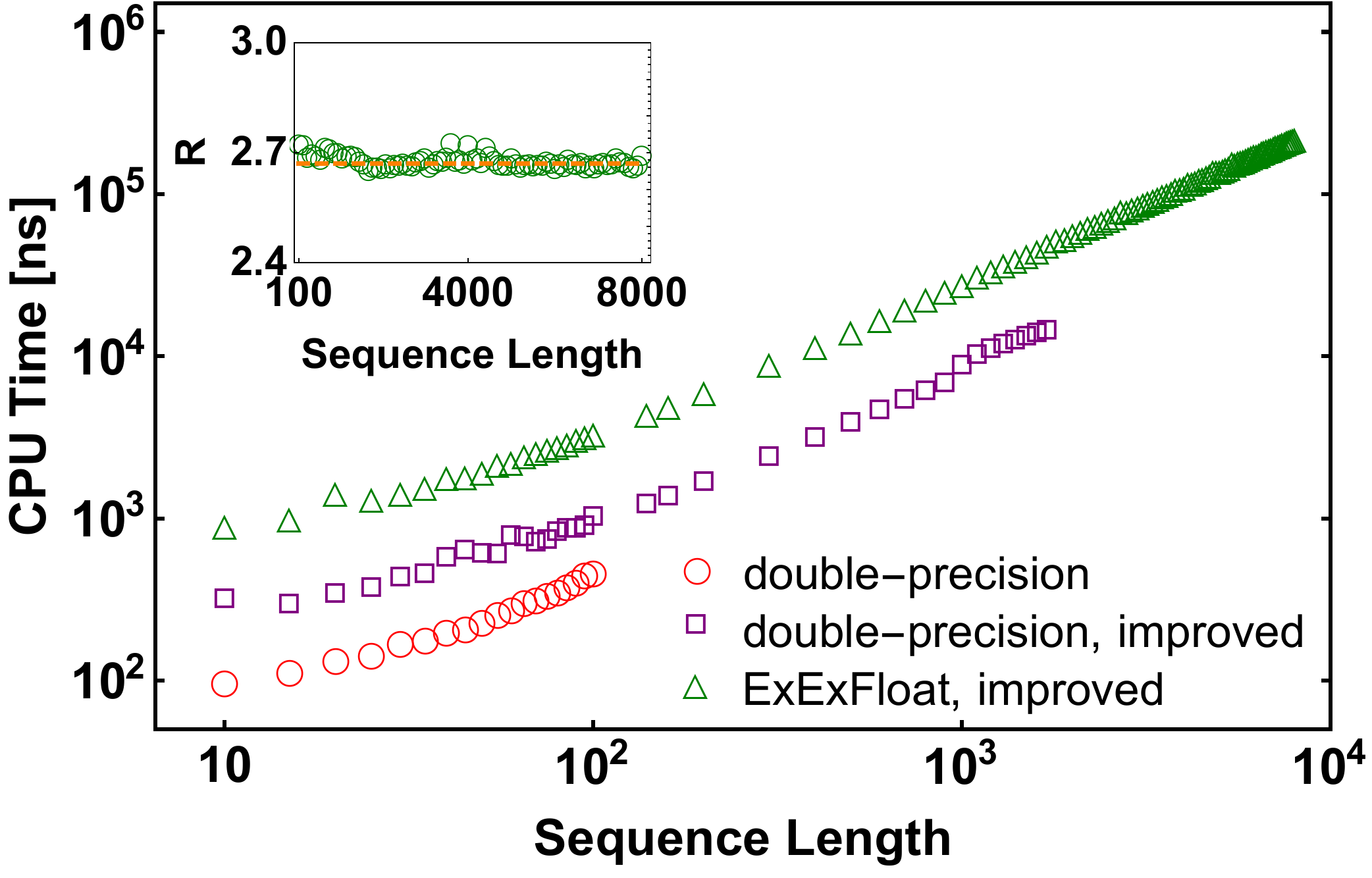}
\small
\cprotect\caption{\label{fig:exexfloat} 
Left: DDEF calculation times with double-precision floating points as a function of $n$, for randomly generated sequences with $s=1$. While the standard formulation breaks down at $n \approx 100$ (circles), the improved formulation allows for an accurate calculation of modified DDEFs of much larger inputs (triangles). 
Right: DDEF calculation times with \verb#ExExFloat# (triangles) vs double-precision floats (squares) as a function of $n$ for $s=2$.
While double-precision floats cannot produce values beyond $n \approx 1700$, calculations with \verb#ExExFloat#  are not hampered. 
Also shown (circles) is the breakdown at $n\approx 100$ of the double-precision calculation of the implementation that does not leverage the improved algorithm formulation discussed in Sec.~\ref{sec:rescale}. In the inset: The ratio $R$ of DDEF calculation time with \verb#ExExFloat# to calculation time using double-precision floating point as a function of input size $n$. The average ratio levels off at $\approx 2.67$ (dashed line). 
} 
\end{figure}

Figure~\ref{fig:exexfloat}(left) shows the scaling of the average DDEF calculation time as a function of input size for input sequences drawn from a random normal distribution with mean $0$ and a standard deviation $\sigma = 0.1$. For these input lists $s=1$, so the scaling step of the algorithm is not executed. Comparing the performance of the improved algorithm against that of the original formulation reveals that while both produce a linear scaling of calculation time with input size as expected, the original algorithm ceases to produce accurate results at around $n \approx 100$ due to an underflow, whereas the improved implementation allows for accurate calculations at much larger sizes.

We also benchmark the performance of the algorithm on input lists whose scaling parameter $s$ is strictly larger than 1 by using inputs sampled from a normal distribution with $\sigma=1$. Here, the scaling step of the algorithm is necessary and a double-precision implementation of the algorithm becomes insufficient beyond certain $n$ and $s$ values. Comparing the performance of an implementation of the addition routine with \verb#ExExFloat# against the double-precision implementation, we make two observations. 
First, we find that
the \verb#ExExFloat# implementation is only about $\approx 2.7$ times slower than its double-precision counterpart (to be compared with the expected cost of arbitrary-precision 
arithmetic which would have resulted in orders of magnitude slowdown~\cite{MPFR}). This is illustrated in the inset of Fig.~\ref{fig:exexfloat}(right),
which shows the ratio of the two computation times for different input sizes. Second, as is evident from the main panel of Fig.~\ref{fig:exexfloat}(right), which depicts the scaling of DDEF calculation time with input size, the improved algorithm with double-precision floats breaks down at $n \approx 1700$. Without rescaling, the breakdown for $s=2$ is observed at $n \approx 100$.
Employing \verb#ExExFloat# on the other hand allows calculating DDEFs of practically unlimited input sizes, going well beyond the capabilities of the current state-of-the-art.

\subsection{Runtimes of the addition and removal routines}

Next, we present the results of several benchmarking tests set out to validate our analysis of the computational complexity of the addition and removal algorithms devised in Sec.~\ref{sec:method}.

\begin{figure}[ht]
\includegraphics[width=0.32\textwidth]{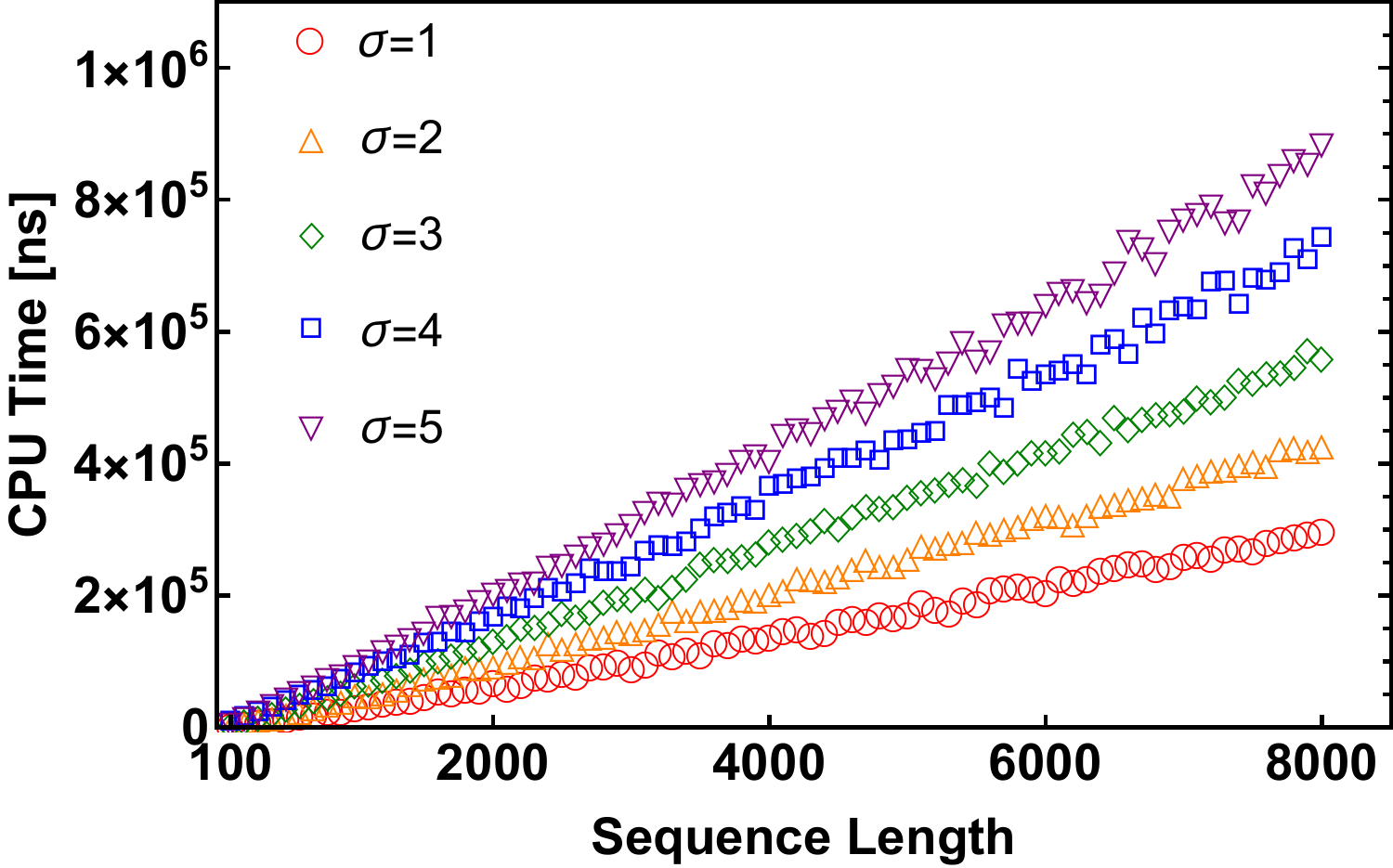}
\includegraphics[width=0.32\textwidth]{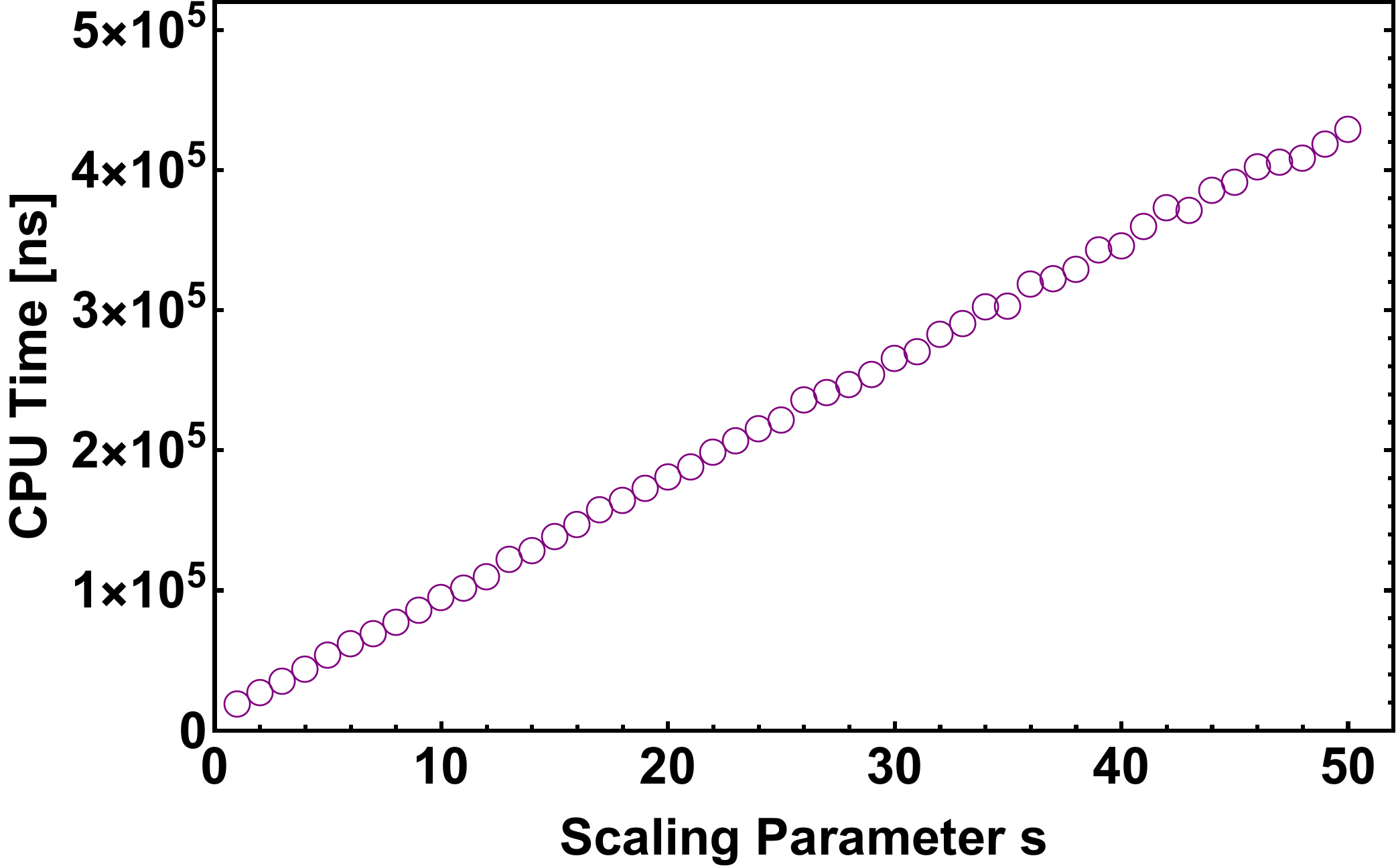}
\includegraphics[width=0.32\textwidth]{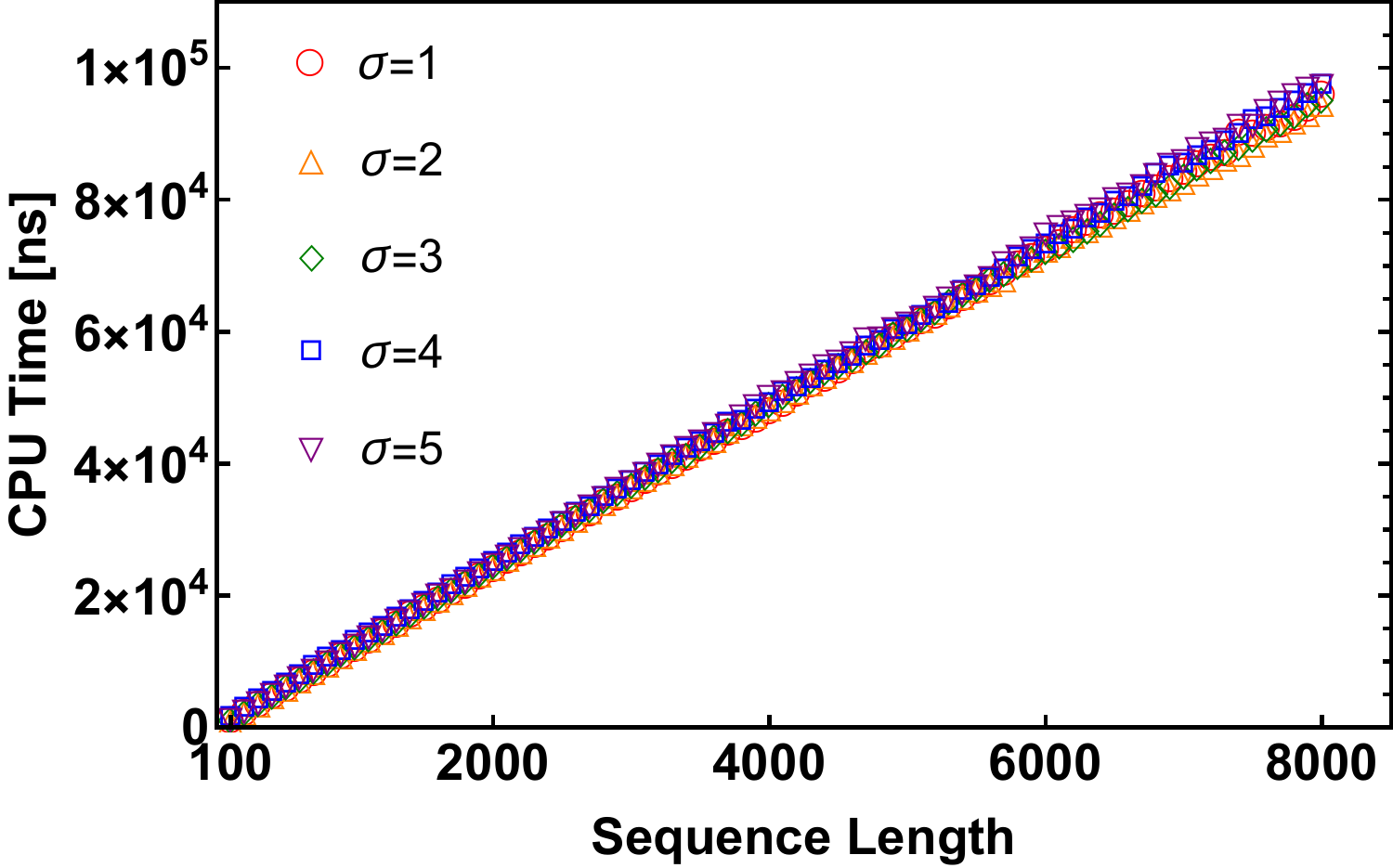}
\small
\caption{\label{fig:bench:add} 
Left: Linear runtime of the addition routine as a function of input size for several different values of $s$. 
Middle: Linear runtime of the addition routine as a function of $s$ for a fixed sequence length ($n=1000$). Right: Linear runtime of the removal routine as a function of input size $n$ for several different values of $s$. The dependence on $n$ is linear whereas the $s$ dependence is marginal. 
}
\end{figure}

Figure~\ref{fig:bench:add} demonstrates the linear runtime scaling of the addition routine with respect to both $n$ (left) and $s$ (middle) confirming the analysis carried out in Sec.~\ref{sec:add}.
In Fig.~\ref{fig:bench:add}(right) we verify that the runtime of removing an item from top of the stack scales as $O(n)$ and has a marginal dependence on $s$,
in accordance with the analysis of Sec.~\ref{sec:remove}.

\section{Applicability to off-diagonal expansion quantum Monte Carlo}
\label{sec:QMC}

The fast and accurate evaluation of DDEFs via addition and removal of inputs has one immediate application in Monte Carlo simulations of quantum many-body systems, specifically in the calculation of configuration weights in off-diagonal expansion (ODE) quantum Monte Carlo~\cite{ODE,PMR}. We discuss this next. 


In statistical physics, a system in thermal equilibrium is completely described by its \emph{partition function}. From it, all thermal properties of the system may be extracted~\cite{Reichl}. For quantum systems, the partition function (denoted $Z$ here) is given by  $Z=\tr \left[ \e^{-\beta H}\right]$
where $H$, the Hamiltonian of the system, is a hermitian matrix, $\tr[\cdot]$ denotes the trace operation
and $\beta$ is a real-valued positive parameter often referred to as the inverse-temperature. For physical systems containing a large number of interacting particles (normally above two dozen or so), the dimension of $H$, which grows exponentially with the number of particles, prohibits the analytical or even numerical evaluation of $Z$, as the calculation requires exponentiating $H$.
In such cases, one often resorts to approximation schemes, usually stochastic methods, within which $Z$ is not evaluated exactly but is rather randomly sampled. This class of techniques is known as quantum Monte Carlo~\cite{newman,Landau:2005:GMC:1051461}. 

Sampling the partition function of a quantum many-body system involves breaking it up to a sum of positive-valued terms 
\beq
Z=\tr \left[ \e^{-\beta H}\right] = \sum_{ \cal C} W_{ \cal C} \,,
\eeq
where each summand $W_{ \cal C}$ is called the `weight' of configuration ${ \cal C}$~\cite{elucidating}.
To sample $Z$, QMC prescribes a Markov process in which a Markov chain of configurations is formed. Starting with a random initial configuration, subsequent configurations are visited with probabilities that are proportional to their weights. For $Z$ to be evaluated properly, the decision of whether to hop from one configuration ${\cal C}_{\text{A}}$ to another ${\cal C}_{\text{B}}$ involves calculating the ratio of their weights, namely, 
$W_{{\cal C}_{\text{B}}}/W_{{\cal C}_{\text{A}}}$.  

Within the ODE quantum Monte Carlo scheme~\cite{ODE,PMR}, a configuration weight is proportional to DDEF
$\exp[z_0,\ldots,z_n]$ whose inputs $z_0, \ldots, z_n$ can be viewed as drawn from a probability distribution over a subset of the diagonal elements of the matrix \hbox{$-\beta H$}. 
The composition and size of the subset  as well as the probability associated with each element are determined by the properties of the Hamiltonian matrix $H$ as well as by the inverse-temperature $\beta$ and so the typical input size $n$ and the scaling parameter $s$ that determine the computational cost of the DDEF (re-)calculation are therefore also strongly dependent on these.
In general, the size of a typical input list is known to scale linearly with the norm of $H$ and the inverse temperature $\beta$ (for more details, the reader is referred to Refs.~\cite{ODE,PMR}). 

Consecutive configurations in the Monte Carlo Markov chain generically have input stacks that differ by $O(1)$ elements so the input stack of any visited configuration can be obtained from that of its predecessor by the addition or removal of a finite number of elements (usually one or two). Since at any given stage of the Markov process the DDEF of the current configuration is already known, a need arises for the fast calculation of the DDEF of the new configuration given that the DDEF of the current one has already been obtained. Since the input stacks of the current and new configurations are very similar, the addition and removal of elements from the input list, followed by the re-evaluation of the DDEF, are a natural way to compute the new configuration weight. 

While the addition of an element can be done by simply using the addition routine, the removal of an item is slightly more involved. This is because the removal of an item from the \emph{bulk} of the stack (rather than from the top) may be called for in the general case. Removal from the bulk is achieved by removing items one by one from the top of the stack and adding all but the target item back. 
In the context of QMC simulations, a pertinent question is therefore how many elements are typically removed and then added back to the stack when a removal of item from the bulk is called for. We address this question next. 

We consider the following scenario, which is typical to QMC.  Let there be a probability distribution $P$ over a discrete set of variables $\mathcal{S}=\{w_1,\ldots,w_M\}$ such that each $w_k$ has probability $p(w_k)$ (and $\sum_{k=1}^M p(w_k)=1$). Next consider a stack containing $n+1$ elements
$z_0,\ldots,z_n$ drawn from $P$ (i.e., $z_j \in \mathcal{S}$). We now pick an additional element $w_j$ from $P$ and ask what is the average number of removals required to remove $w_j$ form the stack. 

The probability that the first occurrence $w_j$ is removed after $r$ sequential removals from the top of the stack is
$P_j^{(r)} = p(w_j)(1-p(w_j))^{r-1}$.
Hence, the average number of removals for $w_j$ is
\beq
\label{eq:meanr}
\langle r\rangle_{w_j} = \sum_{r} r P_j^{(r)} = \frac{1}{p(w_j)}.
\eeq
Averaging over all possible items $w_j$ gives
\beq
\label{eq:number_of_removals}
\langle r\rangle = \sum_j p(w_j) \langle r\rangle_{w_j} = \sum_j \frac{p(w_j)}{p(w_j)} = M\,,
\eeq
the number of participating variables. 
For a finite list of length $n+1$, 
Eq.~(\ref{eq:meanr}) is an approximate relation, because the summation should be performed only
up to $r=n+1$. Moreover, only those items $w_j$ that are likely to be found in the list, i.e., those obeying $\langle r\rangle_{w_j} \leq n+1$, or $p(w_j) \geq 1/(n+1)$, should be taken into account in Eq.~(\ref{eq:number_of_removals}).
Assuming that $P$ is approximately a normal distribution with standard deviation $\sigma$, the number of  values satisfying this condition is
$4\sigma(\Delta w)^{-1} \log^{1/2}\left((n+1)/(\sqrt{2\pi}\sigma)\right)$, where $\Delta w$ is the interspacing between adjacent values.
For as long as $n/\sigma$ is not exponentially large, the number of participating values $M$ is proportional to the standard deviation $\sigma$, which in turn is proportional to $s$. 

In order to remove an element, one needs to perform $r$ removals from the stack and $r-1$ additions to the stack, so the computational cost is $O(rsn)$.
Hence, the average computational cost is $\sum_r O(r s n) P_j^{(r)} = O(sn) / p(w_j) = O(\langle r\rangle_{w_j} sn)$.
Similar to Eq.~(\ref{eq:number_of_removals}), this gives $O(s n M) = O(s^2 n)$ after averaging over all items.
We thus conclude that removal from the bulk requires $O(s^2 n)$ operations.

We next present some results obtained from QMC simulations carried out to ascertain the average number of removals for a particular physical model known as the transverse-field Ising model, which describes a system of particles (or spins) interacting magnetically on a graph (in our simulations we focus on one flavor of this model, namely, 3-regular random antiferromagnets~\cite{farhi:12}). The model is of importance in condensed matter physics as well as in the area of quantum information~\cite{Stinchcombe_1973,farhi:12}. For a system of $N_{\text{s}}$ spins, its Hamiltonian is given by
\beq\label{eq:H}
H= A \sum_{\langle i j \rangle} Z_i Z_j -(1-A) \sum_{i=1}^{N_{\text{s}}}  X_i  \,.
\eeq
Here, $A \in [0,1]$ is a real parameter and ${\langle i j \rangle}$ denotes summation over a subset of pairs of spin indices which constitutes the `interaction graph' of the model~\cite{ODE,PMR}. The matrices $X_i$ and $Z_i$ are a standard shorthand notation for the tensor product of $N_{\text{s}}$ $2 \times 2$ matrices $X_i = \mathds{1} \otimes \cdots \otimes \sigma^x_i \otimes \cdots  \otimes \mathds{1}$ and $Z_i = \mathds{1} \otimes \cdots \otimes \sigma^z_i \otimes \cdots  \otimes \mathds{1}$.
The Pauli matrices $\sigma^x_i$ and $\sigma^z_i$ are given, along with the identity, by 
\beq
\mathds{1} = 
\begin{bmatrix}
    1  &  0      \\
    0  &  1      
\end{bmatrix} \;\;\;,\;
\sigma^x_i = 
\begin{bmatrix}
    0  &  1      \\
    1  &  0      
\end{bmatrix} \;\;,\;\;
\sigma^z_i = 
\begin{bmatrix}
    1  &  0      \\
    0  &  -1      
\end{bmatrix} .
\eeq

\begin{figure}[ht]
\includegraphics[width=0.48\textwidth]{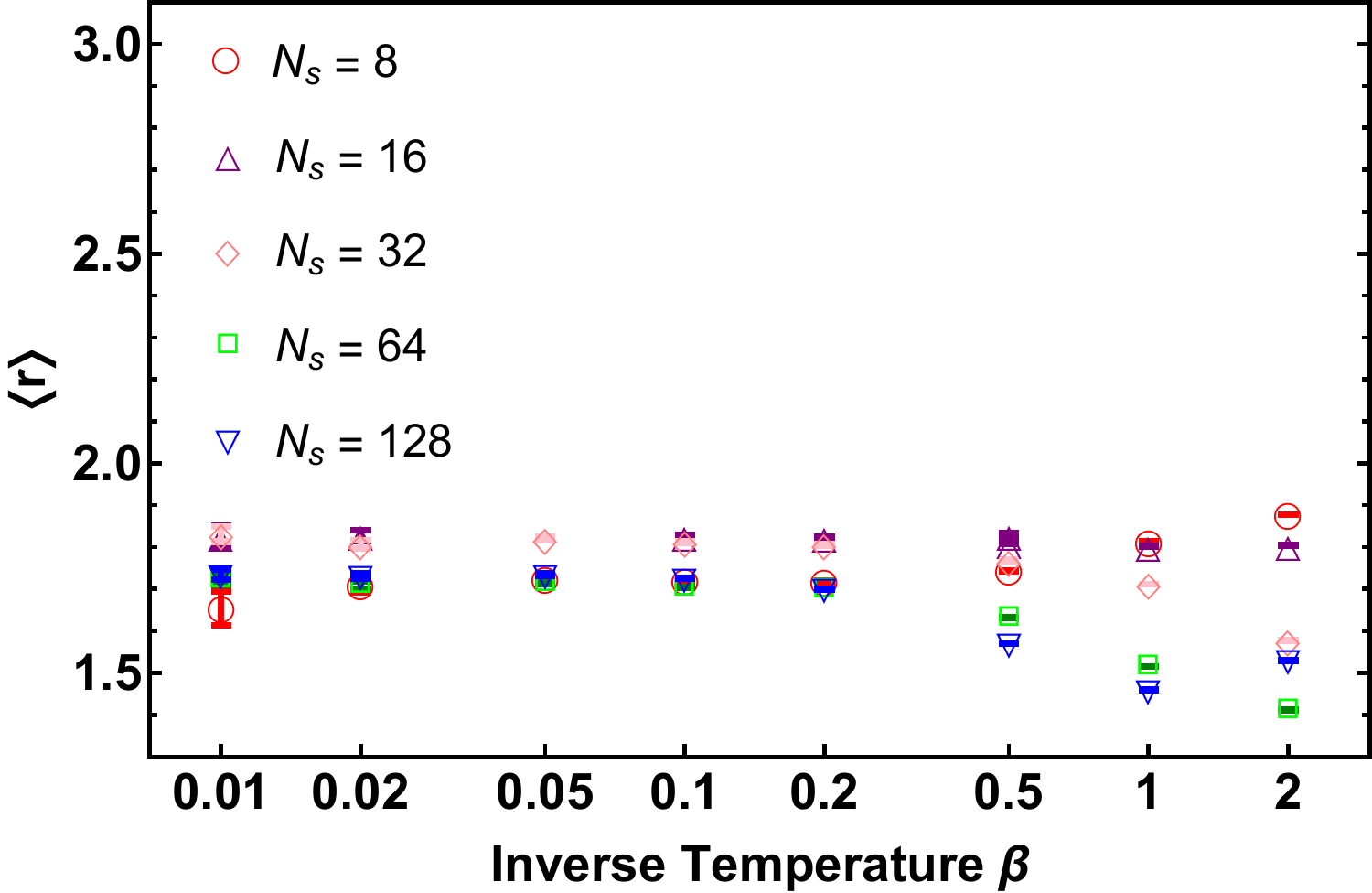}
\includegraphics[width=0.48\textwidth]{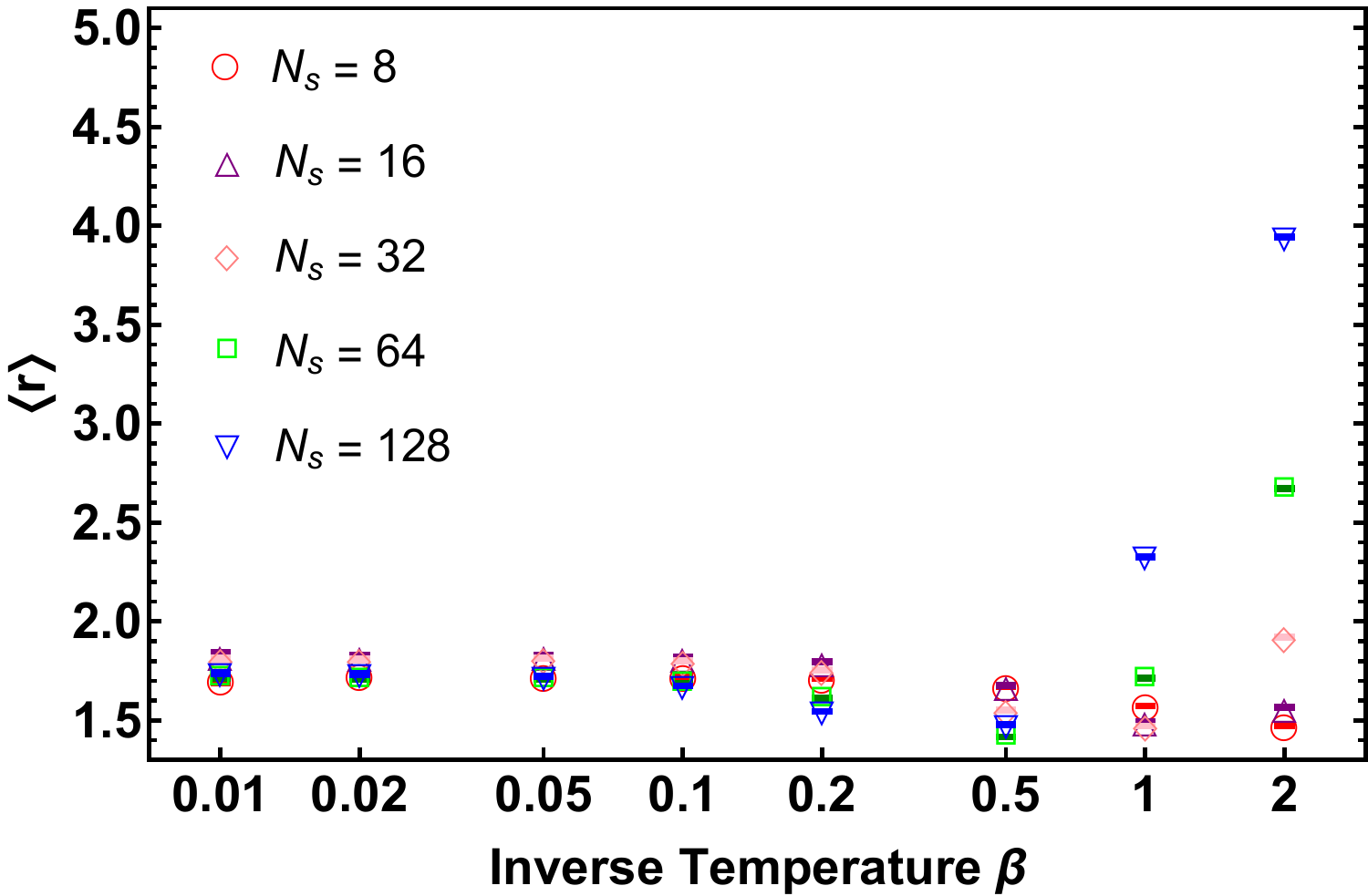}
\small
\caption{\label{fig:avgRemoval} Average number of removals from the top of the stack $\langle r \rangle$ in a single Monte Carlo step as a function of inverse temperature $\beta$ in a QMC simulation of the transverse-field Ising model whose Hamiltonian is given in Eq.~(\ref{eq:H}), for various system sizes $N_{\text{s}}$. Left: $A = 0.2$. Right: $A = 0.5$.  The average stack size $n$ grows linearly with both $N_{\text{s}}$ and $\beta$. Nonetheless, the average number of removals appears to be bounded.} 
\end{figure}

Simulating the above model with ODE QMC, we calculate the average number of removals $\langle r\rangle$ from the top of the stack required within one Markov chain update, as a function of the various parameters of the model.
Figure~\ref{fig:avgRemoval} shows $\langle r \rangle$ as a function of inverse-temperature $\beta$ for $A=0.2$ (left) and $A=0.5$ (right) for different system sizes.

Interestingly, as the figure indicates, the dependence of $\langle r\rangle$ on the inverse temperature is not trivial but is nonetheless bounded across a variety of parameter changes. This behavior may be attributed to the fact that for small values of inverse-temperature $\beta$ the size of the stack is also small (as $\beta \propto n$) and so is the average number of removals. At the other extreme, where  $\beta$ is large, the size of the stack is likewise large. However in this case, the number $M$ of distinct inputs one may find in the stack tends to zero, owing to the fact that DDEFs with only small $z$ values are exponentially more likely to appear than DDEFs containing larger $z$ inputs as the DDEFs serve as configuration weights.

\section{Summary and outlook\label{sec:conclusions}}

We devised an algorithm for calculating the divided differences of the exponential function by means of addition and removal of items from the input list to the function. The addition and removal routines allow the updating of the calculation when the input list undergoes changes. The re-evaluation of the divided differences following the addition or removal of an item from an already evaluated input list requires $O(s n)$ floating point operations and $O(s n)$ bytes of memory, where $n$ is the input size and $s$ is the scaling parameter $\lceil \frac{1}{3.5} \max_{i,j} |z_i - z_j| \rceil$.
Taking advantage of known bounds for divided differences along with a specially devised data structure, our algorithm is able to evaluate the divided differences of the exponential function for input sizes that are orders of magnitude larger than the known capabilities of existing algorithms. 

We also discussed an immediate application of our algorithm in the context of the Monte Carlo simulations of quantum many-body systems, specifically, in the off-diagonal expansion (ODE) QMC algorithm~\cite{ODE,PMR} within which the devised technique can be used to considerably speed up the calculation and extend the range of configuration weights thereby enabling the study of physical models that have so far been inaccessible to ODE. 

We trust that the technique introduced here will enable large-scale simulations of physical models that have so far been beyond the reach of quantum Monte Carlo.
We also hope our method will become useful in other areas where divided differences are needed such as in the calculation of coefficients in the interpolation polynomial in the Newton form and more. We have made our code available on GitHub~\cite{DivDiffCode}.

\begin{acknowledgments}
IH is grateful to Stefan G{\"u}ttel for useful discussions. 
IH is supported by the U.S. Department of Energy, Office of Science, Office of Advanced Scientific Computing Research (ASCR) Quantum Computing Application Teams (QCATS) program, under field work proposal number ERKJ347.
Work by LG is supported by the U.S. Department of Energy (DOE), Office of Science, Basic Energy Sciences (BES) under Award No. DE-SC0020280. 
Work by LB is partially supported by the Office of the Director of National Intelligence (ODNI), Intelligence Advanced
Research Projects Activity (IARPA), via the U.S. Army Research Office
contract W911NF-17-C-0050. The U.S. Government is authorized to reproduce and distribute
reprints for Governmental purposes notwithstanding any copyright notation thereon.
The views and conclusions contained herein are
those of the authors and should not be interpreted as necessarily
representing the official policies or endorsements, either expressed or
implied, of the ODNI, IARPA, or the U.S. Government. 
LB also acknowledges partial support within the framework of State Assignment No. 0033-2019-0007 of Russian Ministry of Science and Higher Education.
\end{acknowledgments}

\bibliography{refs}
\end{document}